\documentclass[aps,prl,twocolumn,superscriptaddress,showpacs]{revtex4}

\usepackage{graphicx}
\usepackage{color}
\usepackage{times}  

\begin{document}
\author{Jessica L.\ McChesney} \affiliation{Advanced Light Source, Lawrence Berkeley National Laboratory, Berkeley, California, USA}
\affiliation{Fritz-Haber-Institut der Max-Planck-Gesellschaft, Berlin, Germany}

\author{Aaron Bostwick} \affiliation{Advanced Light Source, Lawrence Berkeley National Laboratory, Berkeley, California, USA}

\author{Taisuke Ohta} \affiliation{Advanced Light Source, Lawrence Berkeley National Laboratory, Berkeley, California, USA}
\affiliation{Fritz-Haber-Institut der Max-Planck-Gesellschaft, Berlin, Germany}

\author{Konstantin Emtsev}\affiliation{Lehrstuhl f\"{u}r Angewandte Physik, Friedrich-Alexander Universit\"{a}t Erlangen-N\"{u}rnberg, Erlangen, Germany}

\author{Thomas Seyller} \affiliation{Lehrstuhl f\"{u}r Angewandte Physik, Friedrich-Alexander Universit\"{a}t Erlangen-N\"{u}rnberg, Erlangen, Germany}

\author{Karsten Horn} \affiliation{Fritz-Haber-Institut der Max-Planck-Gesellschaft, Berlin, Germany}

\author{Eli Rotenberg} \affiliation{Advanced Light Source, Lawrence Berkeley National Laboratory, Berkeley, California, USA}
\pacs{73.21.-b,71.38.-k,73.22.-f, 79.60.-i}

\title{Self-consistent analysis of electron-phonon coupling parameters of graphene}
\date{\today}

\def\EF{$E_{\mathrm{F}}$}
\def\ED{$E_{\mathrm{D}}$}
\def\TC{$T_{C}$}
\def\bz{Brillouin zone}
\def\htc{high-$T_{C}$}
\def\CaC6{CaC$_{6}$}
\def\C6{C$_{6}$}
\def\KC8{KC$_{8}$}
\def\6r3{$6\sqrt{3}\times 6\sqrt{3}$}

\def\2x2{$(2 \times 2)$}
\def\1x1{$(1 \times 1)$}
\def\RT3{$\sqrt{3}\times\sqrt{3}$-R30}
\def\rt3x{$\sqrt{3}\times$}

\def\kpar{$k_{\parallel}$}
\def\kperp{$k_{\perp}$}
\def\ky{$k_{y}$}
\def\kx{$k_{x}$}
\def\kKpoint{\kpar$=1.703~$\AA$^{-1}$}

\def\A1{\AA$^{-1}$}
\def\vF{$v_{F}$}
\def\kF{$k_{F}$}
\def\me{$m_{e}$}

\def\ee{\emph{e-e}}
\def\eh{\emph{e-h}}
\def\eph{\emph{e-ph}}
\def\epl{\emph{e-pl}}
\def\etal{\emph{et al.}}
\def\pstar{$\pi ^{*}$}
\def\lam{$\lambda$} 

\def\hv{\emph{h}$\nu$}
\def\wb{$\omega_{\mathrm{b}}(\mathrm{\mathbf{k}})$}
\def\wzero{$\omega_{\mathrm{0}}$}
\def\wk{$\omega(\mathrm{\mathbf{k}})$}
\def\wkexpt{$\omega(\mathrm{\mathbf{k}})$}
\def\wkcalc{$\hat{\omega}(\mathrm{\mathbf{k}})$}

\def\kw{$k(\omega)$}
\def\ww{$\mathcal{W}(\omega)$}
\def\wwexpt{$\mathcal{W}(\omega)$}
\def\wwcalc{$\hat{\mathcal{W}}(\omega)$}

\def\wbk{$\omega_{\mathrm{b}}(\mathrm{\mathbf{k}})$}
\def\wltzero{$\omega<0$}
\def\wgtzero{$\omega>0$}
\def\inva{\AA$^{-1}$}

\def\eqakw{Eq.\ \ref{e:akw}}

\def\aw{$\mathcal{A}(\omega)$}
\def\awexpt{$\mathcal{A}(\omega)$}

\def\awcalc{$\hat{\mathcal{A}}(\omega)$}

\def\akw{$A(\mathbf{k},\omega)$}
\def\akexpt{$A(\mathbf{k})$}
\def\akwexpt{$A(\mathbf{k},\omega)$}
\def\akwcalc{$\hat{A}(\mathbf{k},\omega)$}
\def\skw{${\mathrm{\Sigma}(\mathrm{\mathbf{k}},\omega)}$}
\def\iskw{${\mathrm{Im\Sigma}(\mathrm{\mathbf{k}},\omega)}$}
\def\rskw{${\mathrm{Re\Sigma}(\mathrm{\mathbf{k}},\omega)}$}
\def\rskzero{${\mathrm{Re\Sigma}(\mathrm{\mathbf{k}},0)}$}
\def\ims{${\mathrm{Im\Sigma}}$}
\def\res{${\mathrm{Re\Sigma}}$}
\def\t13{$\times10^{13}\ e^{-}/\mathrm{cm}^{2}$}
\def\etal{\textit{et al.}}

\def\mathimskw{\mathrm{Im\Sigma}(\mathrm{\mathbf{k}},\omega)}
\def\mathreskw{\mathrm{Re\Sigma}(\mathrm{\mathbf{k}},\omega)}
\def\mathakw{A(\mathrm{\mathbf{k},\omega})}
\def\slopezero{\partial\mathreskw / \partial k|_{\omega=0}}

\def\akweqn{\begin{equation}\label{e:akw}
\mathakw=\frac{\left|\mathimskw\right|}{\left(\omega-\omega_{\mathrm{b}}(\mathrm{\mathbf{k}})-\mathreskw\right)^{2}+(\mathimskw)^{2}},
 \end{equation}
}

\def\figonecaption{ (a) Experimental and (b) Simulated spectral functions \akw\ for two dopings $n=2.5, 3.9$ \t13.  The dashed lines
are \EF\ and the K point of the \bz, while the solid lines are the self-consistent bare bands.  (c) Real and Imaginary parts of
the self-energy (with 0.02 eV offset), (d) the difference between renormalized and bare bands, and (e) MDC amplitudes for experiment (red,
green) and model (black) as a function of doping.  The lowermost spectra correspond to as-grown graphene ($n=1.2$\t13), with doping
increasing linearly for each subsequent spectrum, up to $n=5.6$\t13.  The heavy lines in (c-e) mark the approximate Dirac energy \ED.}
\def\figone{
\begin{figure*}\centering
{\includegraphics[width=7in]{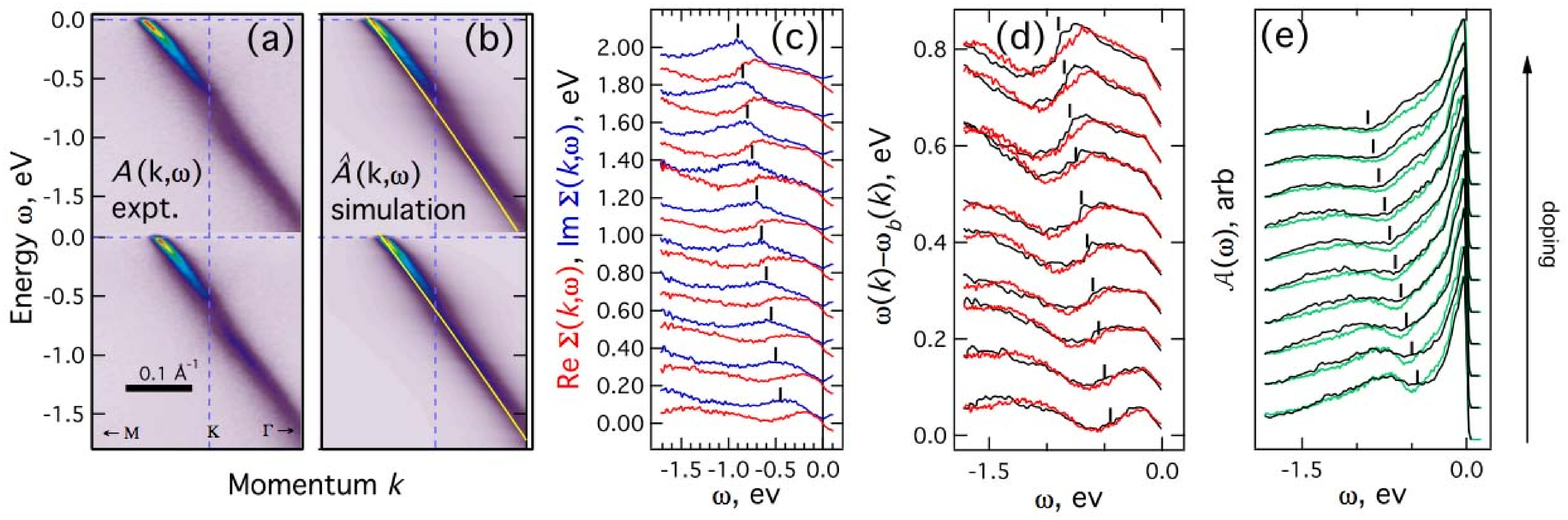} }
\caption{\figonecaption}
\label{fig:1}
\end{figure*}
}


\def\figtwocaption{Simulated spectral function and corresponding self-energy and fitted MDC amplitude \aw\ for 
(a) a single Einstein phonon mode (b) the same, plus an additional Einstein mode at 100 eV of 20\% relative strength, 
(c) a single mode at $\omega_{0}=200$ meV with additional modes decaying towards $\omega=0$.} 

\def\figtwo{\begin{figure}\centering
{\includegraphics[width=3.25in]{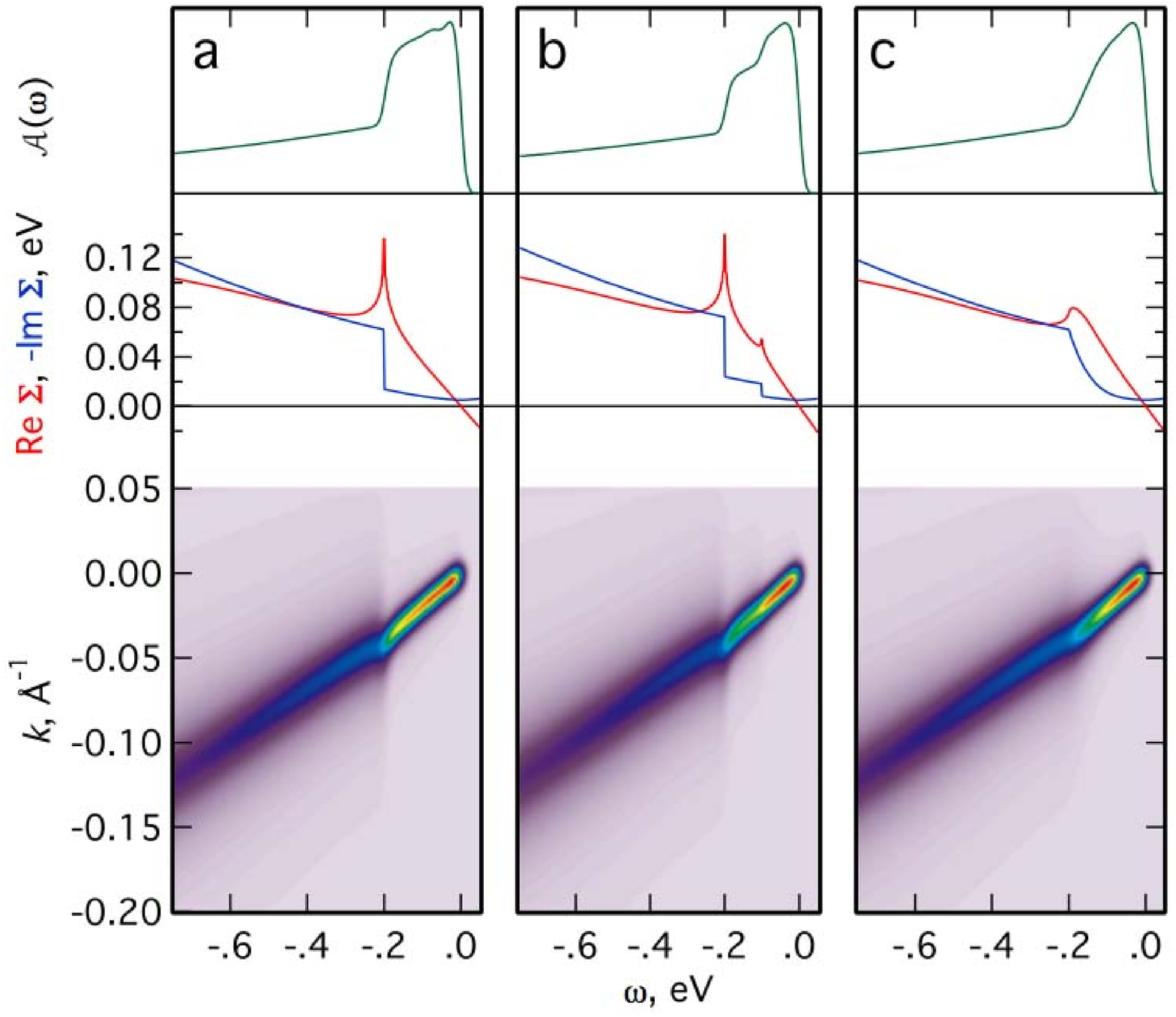} }
\caption{\figtwocaption}
\label{fig:2}
\end{figure}
}

%

\def\figthreecaption{Experimental \aw\ for the same samples as in Fig.\ 1.  The onset position of the phonon kink at $\sim 200$ eV is shown by the dashed line.}

\def\figthree{\begin{figure}\centering
{\includegraphics[width=2in]{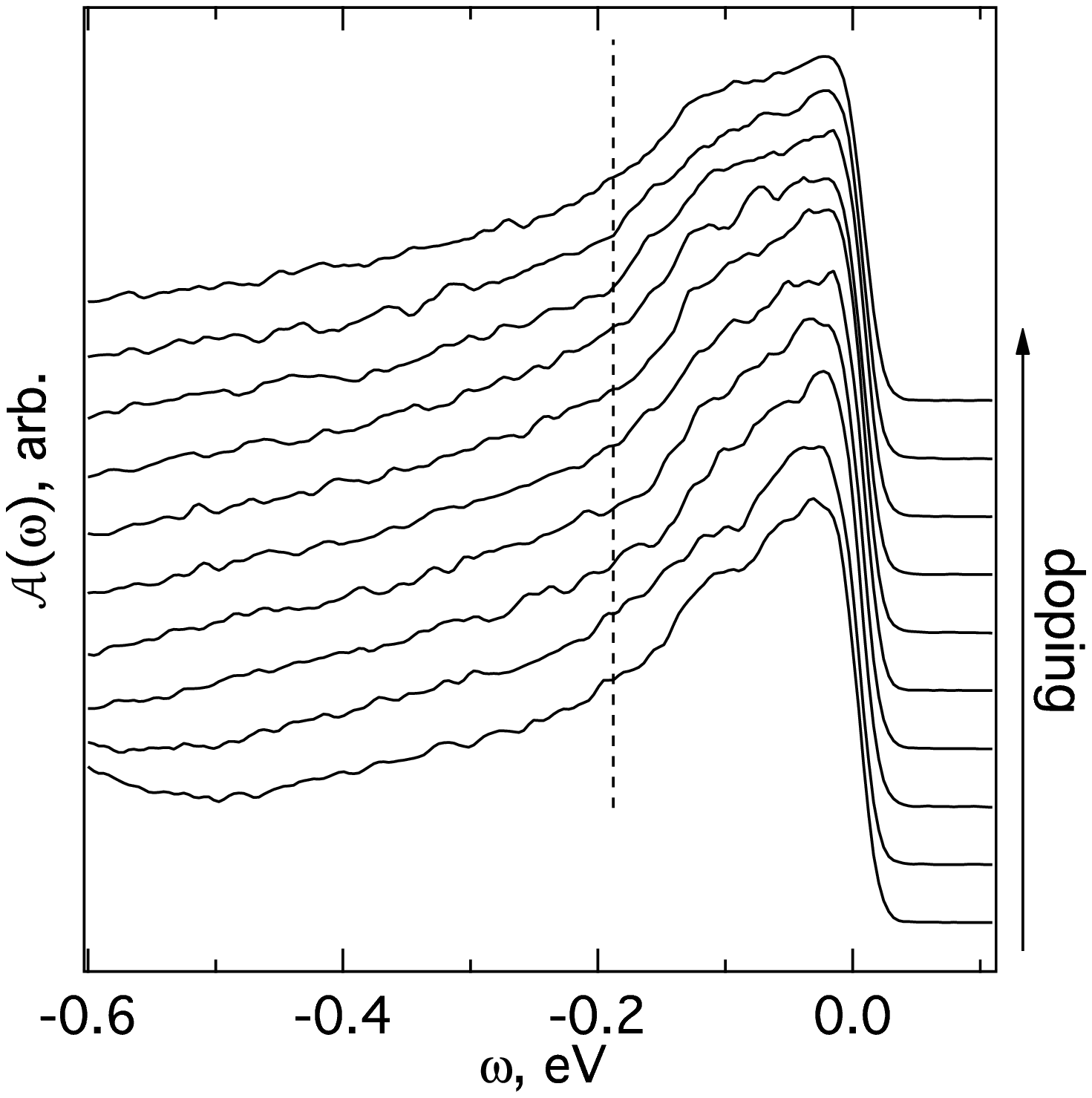} }
\caption{\figthreecaption}
\label{fig:3}
\end{figure}
}

%
\def\figfourcaption{Spectral function of graphene for $n=3.4$\t13 (a) experimental \akwexpt,
(b) optimized model \akwcalc, (c) model \akwcalc\ using self energy by Park \etal\   (d-g) the self-energies, dispersions, MDC widths,
and \aw\ are compared for the results of the MDC analysis of \akwexpt\ (markers) and the models, with (solid, dashed) lines
corresponding to (b, c).}

\def\figfour{\begin{figure*}[t]\centering
{\includegraphics[width=6.5in]{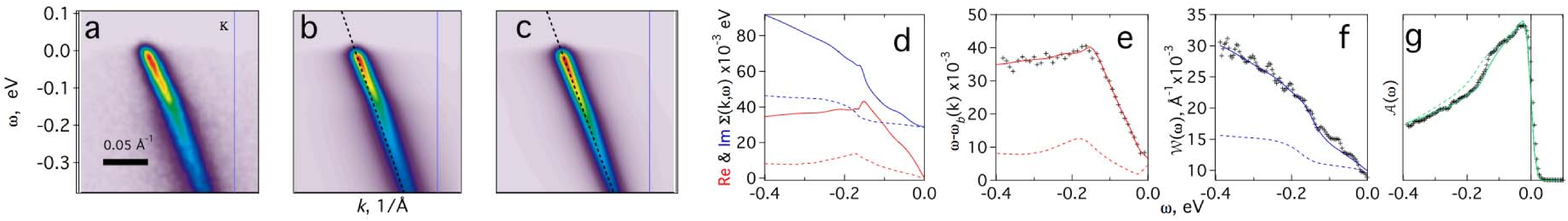} }
\caption{\figfourcaption}
\label{fig:4}       
\end{figure*}
}

\begin{abstract}

We present a self-consistent analysis of the photoemission spectral function \akw\ of graphene monolayers grown epitaxially on
SiC(0001).  New information derived from spectral intensity anomalies (in addition to linewidths and peak positions) confirms that
sizeable kinks in the electronic dispersion at the Dirac energy \ED\ and near the Fermi level \EF\ arise from many-body
interactions, not single-particle effects such as substrate bonding or extra bands.  The relative electron-phonon scattering rate
from phonons at different energy scales evolves with doping.  The electron-phonon coupling strength is extracted and found to be
much larger ($\sim 3.5-5$ times) than predicted.
\end{abstract}
\maketitle
\figone

Many-body interactions in epitaxial graphene are interesting since graphene has been proposed for numerous device applications
including high power electronics and novel device schemes \cite{berger2006,geim2007}.  Full characterization of the electron
scattering lifetime under a variety of conditions is therefore central to understanding the properties of such devices.  Moreover,
graphene is an excellent system for investigating theories of many-body interactions in two dimensions.  First, it is
straightforward to prepare high quality films, with intrinsic photoemission linewidths among the sharpest available for any material
\cite{forbeaux1998,bostwick2007}.  Second, a large change in the carrier density of the order of $\pm10^{13}\ e^{-}/\mathrm{cm}^{2}$
can be achieved through applied voltage in a gated device\cite{novoselov2005,zhang2005}, or equivalently through chemical
doping\cite{ohta2006}, suggesting a tunability of many-body effects for novel devices.

We have previously shown that the electron scattering lifetime of holes in $n$-doped graphene is dominated by a combination of
ordinary Fermi liquid electron-hole pair (\eh) excitations, electron-phonon (\eph) coupling, and electron-plasmon (\epl) coupling
\cite{bostwick2007}.  Theoretical work has qualitatively confirmed the \eh\ and \epl\ interpretation.  Both Hwang \etal\
\cite{hwang2008} and Polini \etal\ \cite{polini2008} have shown that the electron-plasmon interaction leads to an enhancement of the
scattering rate, and a ``$\pi$-band mismatch'', where the lower and upper $\pi$ bands are displaced from each other when projected
through their crossing point at the Dirac energy \ED. These effects are similar to those seen in experiment, but are not predicted
in single-particle theories, whereas they naturally arise from many-body interactions.  Similarly, the dispersion of the bands near
\EF\ is heavily modified, a fact which is ascribed conventionally to \eph\ coupling since the effects occur on a $\sim200$ meV
energy scale, corresponding to the phonon bandwidth of graphene\cite{park2007,tse2007,calandra2007}.

Although the evidence for \eph\ and \epl\ coupling is strong, the existence of the \epl\ coupling and the quantitative estimate of
the \eph\ coupling strength \lam\ derived from the data remain controversial.  First, as an alternate model to the \epl\ coupling,
the bonding of graphene to the substrate induces a gap at \ED \cite{zhou2007,kim2008}.  Alternative causes for a dispersion anomaly
are the presence of defects or quantum size effects for islanded films \cite{rotenberg2008,zhou2008}.

Second, a simple estimate of the coupling strength\cite{bostwick2007b} gave \lam$\sim5$ times stronger than detailed
calculations\cite{calandra2005,calandra2006,park2007,tse2007}.  This suggests a stronger role than
predicted\cite{calandra2005,kim2006} for $\pi$-band \eph\ coupling in superconductivity of graphite intercalation compounds (GICs)
like \CaC6.  But the linear-band estimation method \cite{bostwick2007b} was shown to overstate the coupling\cite{park2008}, and
furthermore finite energy and momentum resolution was cited to explain at least part of the discrepancy \cite{calandra2007}.

Here we provide a self-consistent analysis of the experimental spectral function which makes no \emph{a priori} assumptions about
the bare bands, but assumes only non-violation of causality, approximate particle-hole symmetry, and a weak momentum- (\textbf{k}-)
dependence of the self energy.  Self-consistency between model and data is demonstrated by a comparison of the scattering rates,
dispersion energies, and the \emph{absolute spectral intensity} -- the latter having been not usually considered.  From our analysis
of the low doping regime ($n<6$\t13), we find that: (i) key spectral features arise from many-body interactions, (ii) previous
estimates of large \eph\ constant are confirmed, and (iii) the relative \eph\ coupling strength to multiple phonon modes is shown to
be strongly doping-dependent.  These results are a challenge to the present understanding of the carrier lifetimes in graphene.

The photoemission single particle spectral function is \akweqn where $\omega$ is the quasiparticle energy, \wb\ is the bare
(unrenormalized) band, and \skw\ is the complex quasiparticle self-energy, whose real and imaginary components are related by
Hilbert trasformation to satisfy causality.  We seek to determine \skw\ with no knowledge of the bare band \wb\ or the \wgtzero\
spectral function, and subject to uncertainty due to experimental broadening.  Once \rskw\ is determined, the \eph\ coupling
constant \lam\ is given by \begin{equation}\label{e:lam}\lambda=-\slopezero.\end{equation}\def\eqlam{Eq.\ \ref{e:lam}}

In the usual analysis\cite{damascelli2003,kaminski2005}, each momentum distribution curve (MDC)
$A(\mathrm{\mathbf{k}},\omega$=const$)$ is fitted to a Lorentzian and \akwexpt\ is thereby parametrized into three functions: the
renormalized band dispersion \wk, the Lorentzian width \ww, and \aw, the amplitude along \wk.  Provided the bare band is linear,
\iskw\ is simply proportional to \ww\ and the self-energy is easily found.  However, if \wb\ is non-linear the problem is more
difficult because of the non-trivial relationship between \ww\ and \iskw.

To solve the general problem, we apply an optimization approach similar to Kordyuk \etal\ \cite{kordyuk2005}: From initial guesses
for \wbk\ and \skw, we calculate the simulated spectral function \akwcalc\ and by MDC analysis parametrize it into three functions
\wwcalc, \awcalc, \wkcalc.  The self energy and bare band are iteratively refined in order to minimize the difference
\wkcalc$-$\wkexpt\ subject to the constraint \wwcalc=\wwexpt.  Agreement between simulated and experimental \aw\ is left as a final
check of the model.  To reduce the number of free parameters, we use a quadratic bare band.

The experiments were performed on \emph{in situ} grown samples at beamline 7.0 of the Advanced Light Source.  Samples were prepared
by epitaxial growth on SiC(0001) and doped with K atoms as described elsewhere \cite{ohta2006,bostwick2007}.  The instrumental
broadenings were 25 meV and 0.01 \inva, and the photon energy was 94 eV. The sample temperature was $\sim$20 K.

Fig.\ 1(a,b) shows two typical experimental spectral functions \akwexpt\ for different $n$-dopings of graphene, in comparison to
simulated spectral functions \akwcalc.  The latter were calculated using the optimized bare bands and \skw\ shown in Fig.\ 1(b,c).
We assumed particle-hole symmetry, i.e. that \iskw\ is a strictly even function with respect to $\omega=0$, implying
\rskzero$\rightarrow0$.  The intensity was scaled by a function linear with $\omega$ to account for the non-uniform sensitivity of
the electron detector.  Comparison of the quantity (\wk-\wbk) derived from MDC fits of \akwexpt\ and \akwcalc\ (Fig.\ 1(d)) shows
excellent agreement, demonstrating self-consistency of our derived \skw, and confirming that deviations in the band dispersion
indeed are fully described by quasiparticle scattering, and not details of the initial-state band structure.  In particular, the
energy gap at \ED\ \cite{zhou2007,kim2008}, if any, must be much smaller than the band renormalization by many-body interactions.

We stress that in our analysis the simulated and experimental spectral functions are treated on an equal footing with respect to
temperature smearing at \EF\ and experimental broadening, both of which are included in our simulated \akwcalc.  Therefore these
uncertainties are in principle deconvolved out of our derived self-energies.  Furthermore, we find that the calculated and
experimental \aw\ in excellent agreement (Fig.\ 1e), despite not being included in our optimization.  This is a stringent test of
the analysis, because as we now show, \aw\ is highly sensitive to many-body interactions, perhaps more so than the MDC width and
peak positions, a fact we can exploit to determine new information about \eph\ coupling.

Fig.\ 2(a) shows simulated spectral functions for a band interacting with a single Einstein optical mode at \wzero=200 meV using
realistic experimental parameters, including a power-law scattering rate to simulate electron-electron interactions.  The MDC
amplitude function \aw\ shows a relatively flat top down to \wzero.  Suppose we add additional weak scatterers at lower energy
scales (Fig.\ 2(b,c)); how is the MDC analysis affected?  Casual inspection of the \akw\ images shows that scarcely any effect on
the apparent dispersion or scattering rates can be observed with realistic broadening and limited statistics.  However, the MDC
amplitudes \aw\ are strongly affected, with an easily visible shift of spectral weight from \wzero\ towards \EF. This shows that
\aw\ can access information about weak or even sub-resolution features in the self-energy.

We believe that the evolution of such `hidden' features with doping are necessary to explain the evolution of \aw\ near \EF. Fig.\ 3
shows the measured \aw\ as a function of doping.  While all spectra show a break in slope around 200 meV, there is clearly an
evolution from a merely prominent shoulder beginning around 200 meV to a flat top below \EF\ with increased doping.  The simulations
in Fig.\ 2 suggest that this shoulder is the dominant mode at all dopings, but its strength increases relative to the lower energy
modes with doping.  Such a relative evolution of the \eph\ coupling with different phonon modes has not been predicted by theory.

\figtwo

We close by discussing how the self-energy and experimentally derived \eph\ coupling constant \lam\ compares to recent calculations.
The globally optimized self-energies in Fig.\ 1 do not fit perfectly well near \EF, since the quadratic bare band does not fit the
data perfectly over all energies.  To achieve a more accurate \skw\ near \EF, we have analyzed the data in a smaller energy region,
shown in Fig.\ 4(a) for a sample with \EF$-$\ED\ $\sim .65$ eV.

\figthree

The simulated \akwcalc\ and the the derived self-energy \skw\ and bare band \wb\ are shown in Fig.\ 4(b, d).  Good agreement between
measured and simulated \wk, \ww, and \aw\ are demonstrated in Fig.\ 4(e-g).  For comparison, in Fig.\ 4(c) we show the spectral
function calculated from the predicted \eph\ \skw\ from Ref.\ \cite{park2007}.  Both the experimental and the predicted \akw\ were
modeled with a .028 eV offset to \iskw\ to take care of the background defect scattering rate.  Although in excellent qualitative
agreement, the theoretical self-energies are much smaller than those measured (see comparison in Fig.\ 4(d)), and hence the kink,
scattering rates, and the abrupt increase in the spectral amplitude \aw\ are not well-reproduced (Fig.\ 4(e-g)).  This conclusion is
not affected by experimental broadening as proposed in \cite{calandra2007} because the same discrepancy appears in both the
optimized self-energies (from which the broadening is essentially deconvolved, see Fig.\ 4(d)) as well as from the apparent
self-energy derived from both experiment and broadened theory (Fig.  4(e-f)).  Furthermore the disagreement in \aw\ is practically
unaffected by our experimental broadening.  \figfour

The discrepancy in \lam\ ranges from a factor of 3.5 to 5 times larger than predictions for \eph\
coupling\cite{calandra2007,park2007,tse2007}, the lower estimate found by scaling the predicted real and imaginary parts of \skw\ to
match the experimental functions, and the upper estimate by applying \eqlam\ to the two \rskw\ functions in Fig.\ 4(d) \cite{note2}.
If some of the scattering is attributed to electron-electron coupling near \EF\ (which rises as $\sim\omega^{2}$) \cite{polini2008}
then this would act to reduce the derived \eph\ coupling constant, however, the \ee\ contribution to \iskw\ is predicted to be very
weak (around .004 eV at $\omega=200$ meV) \cite{note1}, and it cannot explain the sharp kink at the phonon energy scale
\cite{polini2008,hwang2008}.

Why is the apparent \eph\ coupling so large?  First, it could be due to coupling to some other mode (e.g. magnon) at similar energy
scale, although such a mode has not been observed or even predicted.  It cannot be due to coupling to substrate phonons, because the
SiC phonon modes are at lower energies than the observed kink.  Furthermore there is a carbon-rich, graphene-like buffer layer
between graphene and SiC \cite{emtsev2007}, whose coupling to the graphene electrons would also have to be anomalously high to
explain the results.

The presence of our dopant K atoms cannot explain the large coupling: first, because of the apparent strong coupling even in our
as-grown films, and second, because K vibrations are at too low an energy to explain the kink (although they might contribute in
principle to the lower-energy modes suggested in our self-energy in Fig.\ 4(d)).  Also, the very minute amount of K atoms ($\sim.06$
per graphene unit cell at our highest doping) makes its presence unlikely to cause the effect we see.  Lastly, there is a small
rippling of graphene on SiC \cite{varchon2008}.  Although a local curvature can enhance the \eph\ coupling \cite{crespi1999}, the
curvature in our films is too small to have a significant effect.

Our findings lead to three possibilities, first, that $\lambda$ for graphene is anomalously strong, second, that there is
anomalously strong scattering from something other than phonons (such as defects, or \ee\ scattering) on a similar energy scale, or
third, there is a mutual interplay of interactions that leads to an enhanced scattering rate overall.  Noting that the peak near
\ED\ in \ims\ due to plasmon scattering is broadened and shifted towards \EF\ in our experiment relative to calculations
\cite{polini2008,hwang2008}, such an interplay is conceivable.

We acknowledge C. W. Park, F. Guistino, S. G. Louie, M. L. Cohen, M. Polini, A. H. Macdonald, T. Deveraux, M. Calandra and F. Mauri
for discussions.  This work and the ALS are supported by the Director, Office of Science, Office of Basic Energy Sciences, Materials
Sciences Division, of the U.S. DOE under Contract No.\ DE-AC03-76SF00098 at LBNL. J.M., T.O., and K.H. were supported by the Max
Planck Society.  TS and KE acknowledge support by the DFG through SE1087/5-1 and by the Erlangen Cluster of Excellence "Engineering
of Advanced Materials".  (http://www.eam.uni-erlangen.de).


\begin{thebibliography}{29}
\bibitem[{not({\natexlab{b}})}]{note1}
\bibinfo{note}{The quasiparticle peak at $\omega$=0.2 eV corresponds to $\omega
  /(E_{\mathrm{F}}-E_{\mathrm{D}})=0.3$, and normalized momentum
  $k/k_{\mathrm{F}}\sim 0.75$. The predicted
  Im$\mathrm{\Sigma}/(E_{\mathrm{F}}-E_{\mathrm{D}})\sim .0059$ is read from
  Fig.\ 4 of Ref.\ [9], or Im$\mathrm{\Sigma}\sim .0038$ eV}.
\end{thebibliography}

\end{document}